# Decomposition of the absorbed dose by LET in tissue-equivalent materials within the SHIELD-HIT transport code

N. Sobolevsky[1,2], A. Botvina[1], N. Buyukcizmeci[3], A. Ergun[3], L. Latysheva[1], R. Ogul[3]
[1] Institute for Nuclear Research RAS, 117312 Moscow, Russia
[2] Moscow Institute of Physics and Technology (MIPT), 141700 Dolgoprudny, Russia
[3] Selcuk University, Department of Physics, 42079 Konya, Turkey

**Abstract**
The SHIELD-HIT transport code, in several versions, has been used for modeling the interaction of therapeutic beams of light nuclei with tissue-equivalent materials for a long time. All versions of the code include useful option of decomposition of the absorbed dose by the linear energy transfer (LET), but this option has not been described and published so far. In this work the procedure of decomposition of the absorbed dose by LET is described and illustrated by using the decomposition of the Bragg curve in water phantom, irradiated by beams of protons, alpha particles, and of ions lithium, carbon and oxygen.

**1. Introduction**

Biological effects of irradiation in hadron therapy depends not only on the amount of energy absorbed per mass unit (i.e., on the absorbed dose D) but also on the equivalent absorbed radiation dose (the equivalent dose H), which essentially depends on the LET by hadrons and nuclear fragments in irradiated tissue.

At high LET, the equivalent dose H measured in sieverts (Sv) is substantially higher than the absorbed dose (physical) D measured in grays (Gy). The relation is given by

$$H = D \times K,$$

where K is a dimensionless quality factor of the radiation ($1 < K < 20$). It follows that 1 Sv=1 Gy/K, since equivalent dose of H=1 Sv is dialed when the absorbed dose D =1 Gy/K.

It is therefore important to know not only the amount of energy released in a given volume of a target, but also values of LET at deposition of this energy. Thus, the problem of decomposition of the absorbed dose by LET arises, i.e. in what intervals of LET the energy release occurs.

The SHIELD-HIT transport code (http://www.inr.ru/shield/) has been repeatedly used for modeling of interaction of therapeutic beams of hadrons and light nuclei with tissue-equivalent materials (see, for example, [1-7]). In this paper one more option of the code - the possibility of decomposition of the absorbed dose by LET is presented for the first time. As an illustration, the decomposition by LET of the Bragg curve for cylindrical water target which is irradiated by protons and ions $^4He^{+2}$, $^7Li^{+3}$, $^{12}C^{+6}$ and $^{16}O^{+8}$ is given by the Figures in the text. Energies of the projectiles correspond to the average range to stop of 26 cm.

## 2. Algorithm of decomposition of the absorbed dose by LET in the SHIELD-HIT code

In the context of hadron therapy, LET coincides with the stopping power dE/dX, i.e. LET ≡ dE/dX, although in principle LET and dE/dX are corresponding to various physical quantities. The designations LET and dE/dX are used further as synonyms.

Fig. 1 shows the data recommended by ICRU [8,9] for the stopping power of water for protons and light nuclei in the energy range of 0.025-1000 MeV/A.

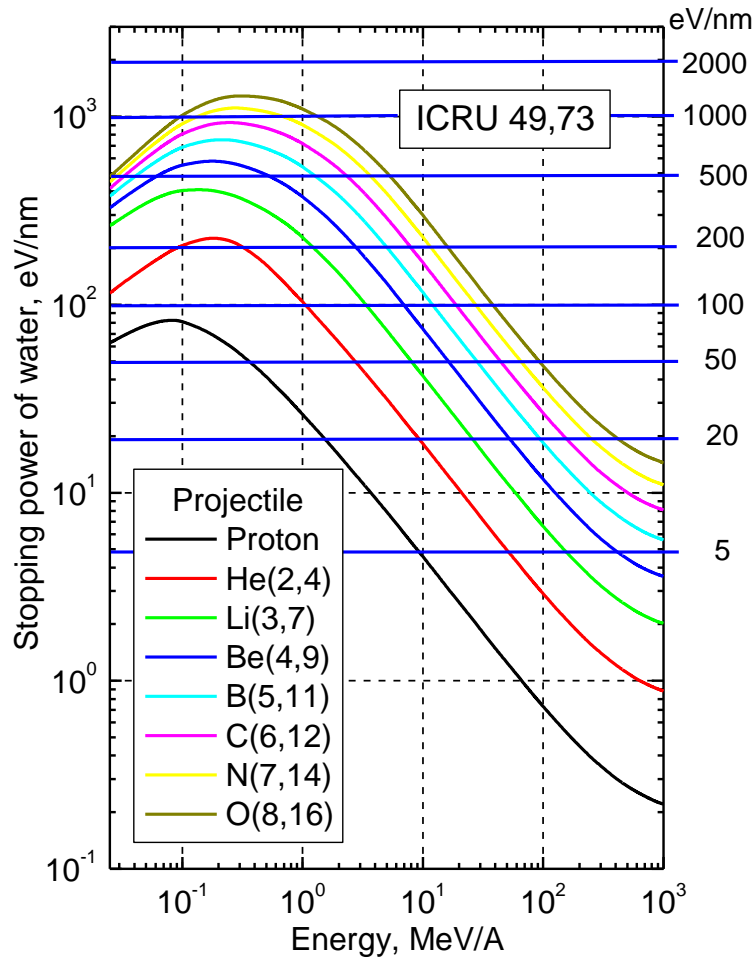

Fig. 1. The stopping power of water for light ions in the therapeutic range of energies.

The user of the SHIELD-HIT code can set the intervals of partition on LET, for example, as shown in Fig. 1. This set of intervals is the same for all primary and secondary particles and nuclear fragments (projectiles) in given task as well as for different materials in different geometrical zones of the target.

Suppose that in the process of simulation a particular projectile crosses a known geometric zone of the target containing a specific substance. In this case the SHIELD-HIT code detects energy intervals ΔE, that match specific intervals of LET on the stopping power curve dE/dX(E), as shown in Fig. 2. The borders of the detected intervals ΔE are memorized in a special array.

Since a dE/dX(E) curve includes the increasing and decreasing parts with energy, for a given interval of LET there may be two energy intervals ΔE1, ΔE2, only one interval ΔE, or no intervals ΔE, if the curve dE/dX(E) lies below the predetermined interval of LET, see. Fig. 2.

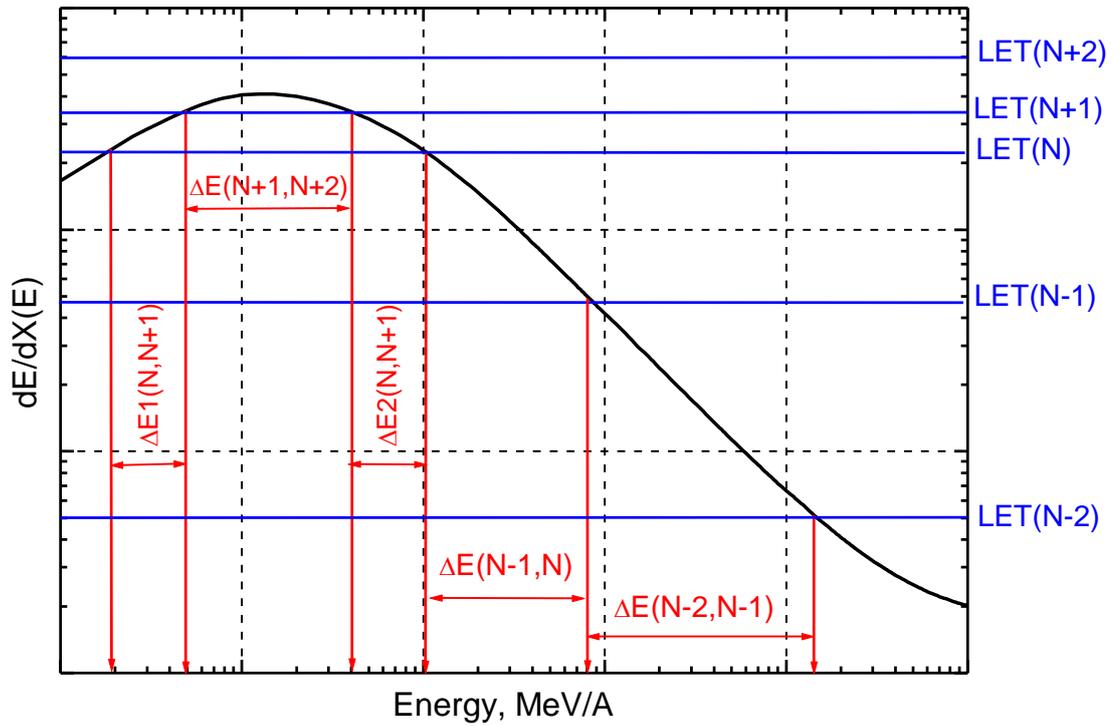

Fig. 2. Layout of LET and corresponding energy intervals ΔE on the curve dE/dX(E).

Further steps of the algorithm for decomposition of the dose by LET depend on the relative location of boundaries of the interval ΔE, and of initial energy $E_{init}$ and final energy $E_{finl}$ of the projectile as it passes through the zone. The initial energy $E_{init}$ corresponds to the entry point of a projectile into the zone or its birth point within the zone. The final energy $E_{finl}$ corresponds to the exit point of a projectile out of the zone or to the point of its absorption/interaction within the zone.

In addition, the procedure for decomposition of the dose by LET depends on the position of points $E_{finl}$ and $E_{init}$ under the curve dE/dX(E). The case where the initial energy $E_{init}$ is located under the decreasing part of the curve dE/dX(E), while the final energy $E_{finl}$ - under the increasing part of this curve, is shown in Fig. 3. In the figure, the values of contributions to the energy deposition ΔQ2 and ΔQ1 depending on the position of points $E_{init}$ and $E_{finl}$ as well as the total energy deposition in the zone ΔQ=ΔQ1+ΔQ2, provided that the value of LET is in the range [LET(N),LET(N+1)], are shown.

The cases when both values $E_{init}$ and $E_{finl}$ are located either under the decreasing or increasing parts of the curve dE/dX(E), are considered separately. Fig.4 illustrates the first one of these two cases. The energy deposition ΔQ2 is different from zero only if the intervals [$E_{finl}$,$E_{init}$] and [E3,E4] overlap each other, at least partially. Under this condition the energy deposition ΔQ2 takes one of the values of $ΔQ_{ac}$=E4−$E_{finl}$, $ΔQ_{ad}$=E4−E3, $ΔQ_{bc}$=$E_{init}$−$E_{finl}$, $ΔQ_{bd}$=$E_{init}$−E3.

The second case, when both values, $E_{init}$ and $E_{finl}$, are located under the increasing part of the curve dE/dX(E) is treated similarly. For the interval of LET in which the maximum of the curve dE/dX(E) is disposed, the consideration is carried out separately as well.

Thus, the energy deposition by the projectile in a particular geometric zone of the target on the condition that LET is within a predetermined range values, has been calculated, which means the decomposition of the absorbed dose by LET.

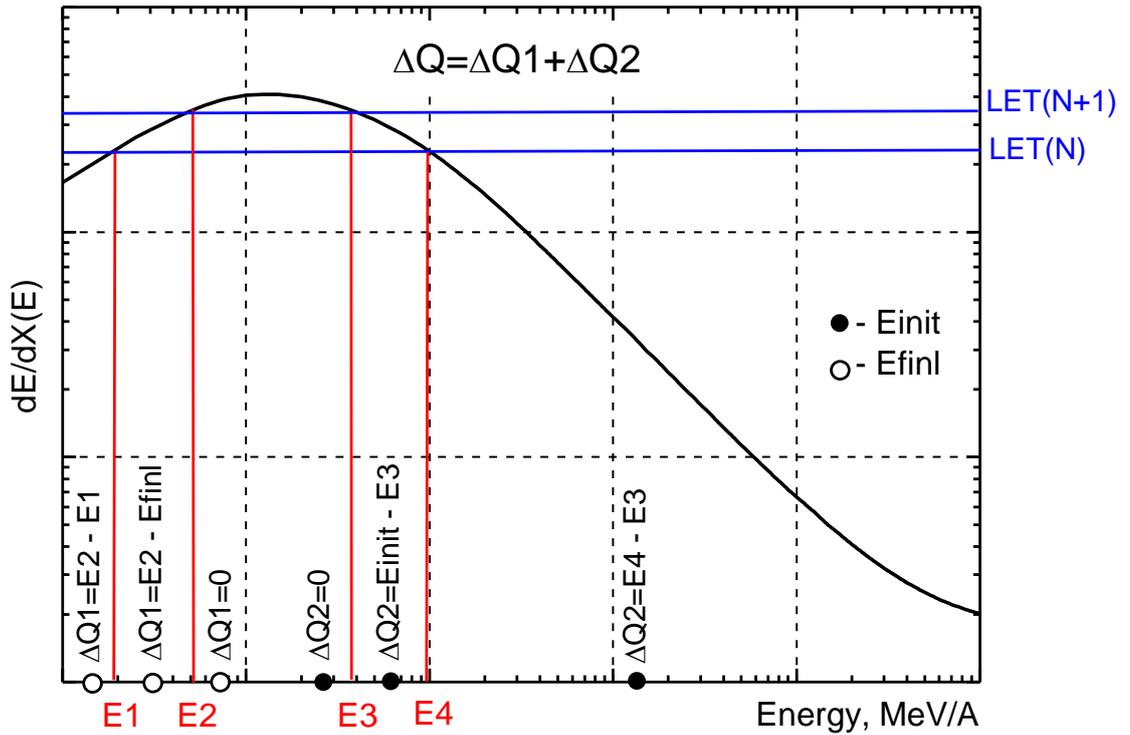

Fig. 3. Layout of points $E_{init}$ and $E_{finl}$ on the energy axis when the point $E_{init}$ is under the decreasing part of the curve dE/dX(E), while $E_{finl}$ is under the increasing part. Formulas for the energy deposition $\Delta Q$ depending on the position of $E_{init}$ and $E_{finl}$ are given.

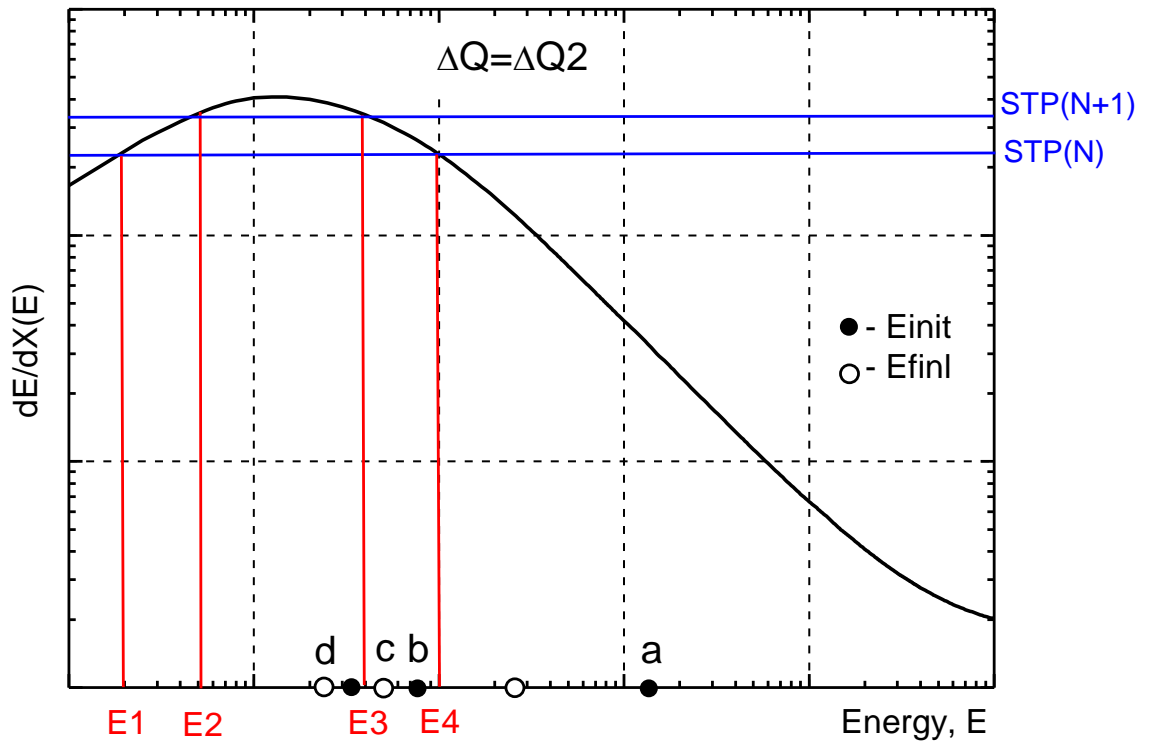

Fig. 4. Layout of points $E_{init}$ and $E_{finl}$ on the energy axis when both points are under the decreasing part of the curve dE/dX(E). Letters a,b,c,d mark the positions of the points that give contribution to the energy deposition $\Delta Q$ (see text).

Finally let us clarify, how the SHIELD-HIT code accumulates the energy deposition (the absorbed dose) in given geometric zone of a target. The following channels of the energy deposition are taken into account:
- Ionisation loss of the transported charged hadrons and nuclear fragments in the geometric zone of a target, from the entrance to the zone (birth within the zone) up to leaving the zone (interaction/absorption inside the zone).
- The energy of a residual nucleus or recoil nucleus at the interaction / scattering in the target zone. This energy can be both above and below the cutoff energy of the projectile during transportation $E_{cut}$=0.025 MeV/A. In the first case the residual nucleus is transported like a nuclear fragment.
- The local energy deposition of transported hadron or nuclear fragment when it reaches the cutoff energy $E_{cut}$ in given geometric zone (residual energy deposition 0.025·A MeV) as well as the energy of a residual/recoil nucleus if it is below $E_{cut}$.

**3. Results**

Figs.5-9 show the results of present simulations for the decomposition of the Bragg curve by LET in water. Narrow beams of protons and ions $^4$He$^{+2}$, $^7$Li$^{+3}$, $^{12}$C$^{+6}$ and $^{16}$O$^{+8}$ impinge on the cylindrical water target along its axis. The water target of radius R=10 cm and of length L=40 cm is divided into layers having the thickness ΔL=0.1 cm, so that there are total 400 geometric zones. Energies of the projectiles correspond to the average range to stop of 26 cm, i.e. are equal to 202 MeV/A for protons and $^4$He$^{+2}$ ions, and 233 MeV/A, 391 MeV/A and 469 MeV/A for the ions $^7$Li$^{+3}$, $^{12}$C$^{+6}$ and $^{16}$O$^{+8}$, respectively. The results are normalized to the number of primary projectiles and to the thickness of the layer ΔL, i.e. are presented in units MeV/(cm·projectile).

In the simulation we took into account all the generations of secondary particles produced in inelastic nuclear interactions, as well as elastic nuclear scattering, fluctuations of ionization losses (according to Vavilov-Landau theory) and multiple Coulomb scattering (Moliere theory). A more detailed description of capabilities of the SHIELD-HIT code can be found in the papers [1-7] cited above.

The SHIELD-HIT code allows us to decompose not only the total energy deposition in the target (the Bragg curve in this case), but also the contribution to the energy deposition of a separate secondary fragment. Fig. 10 illustrates decomposition by LET of contribution of the secondary ion $^{11}$C$^{+6}$ into the Bragg curve at irradiation of the water target by the ion beam $^{12}$C$^{+6}$.

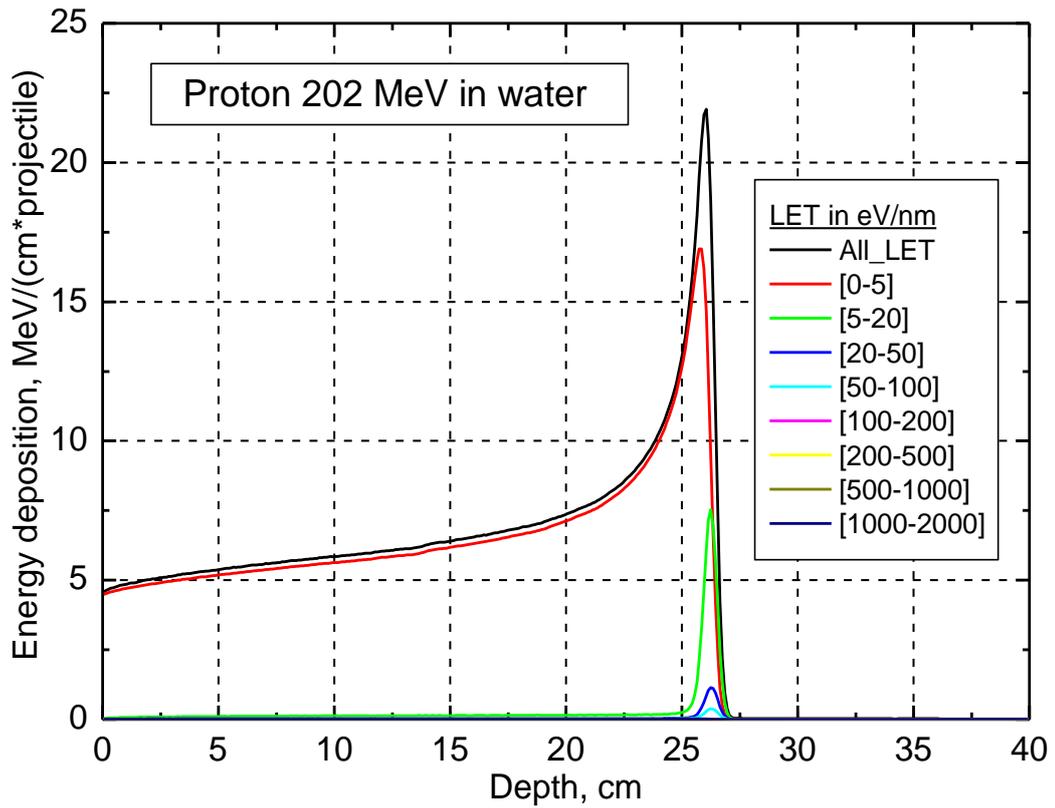

Fig. 5. Decomposition of the Bragg curve by LET in water for projectile–proton at 202 MeV.

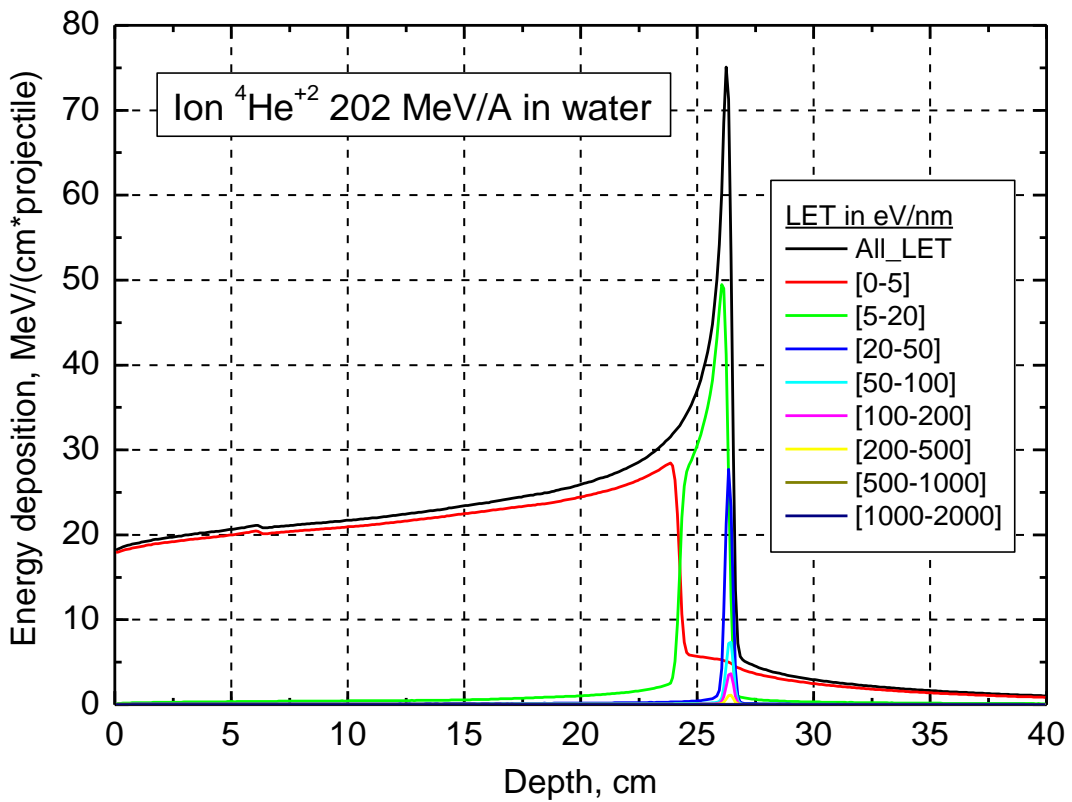

Fig. 6. Decomposition of the Bragg curve by LET in water for the projectile-ion $^4\text{He}^{+2}$, 202 MeV/A.

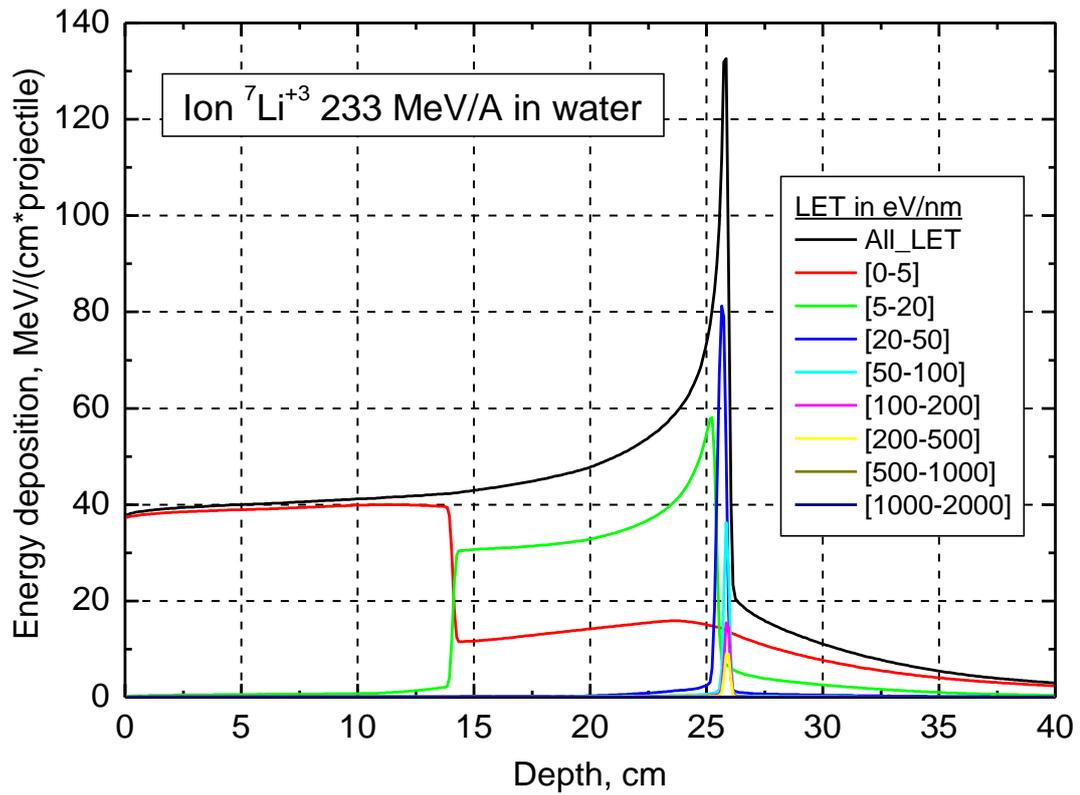

Fig. 7. Decomposition of the Bragg curve by LET in water for the projectile-ion $^7Li^{+3}$, 233 MeV/A.

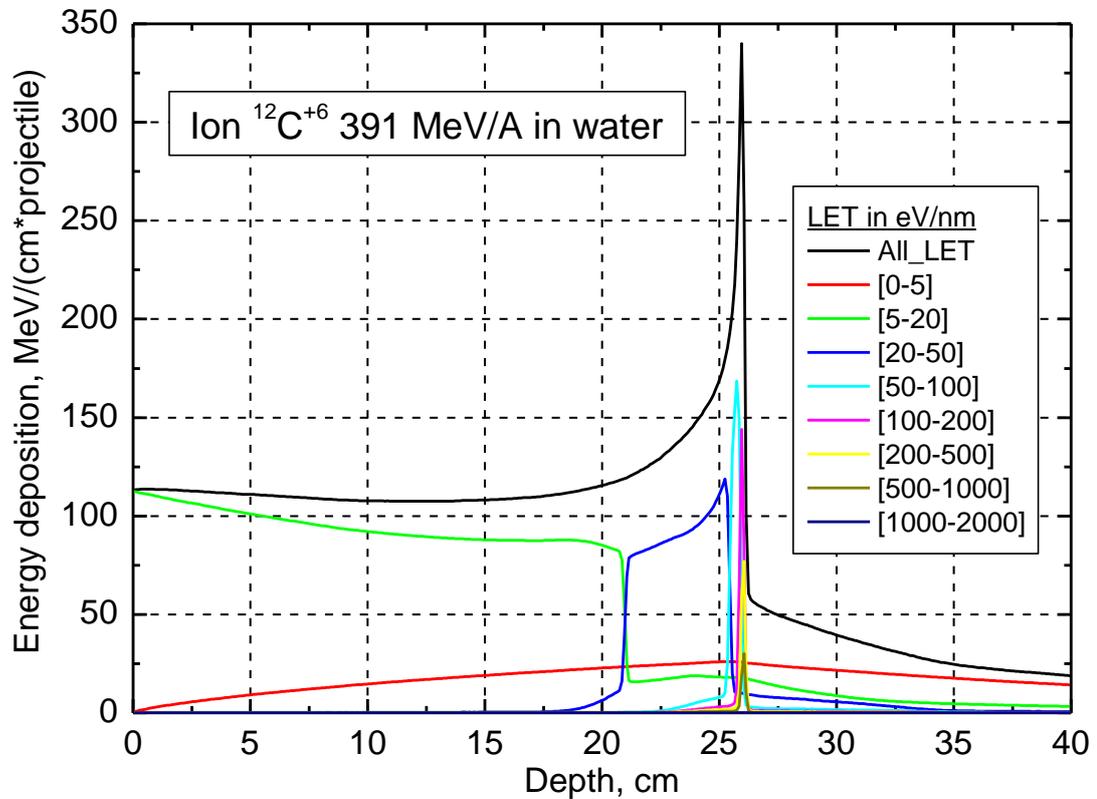

Fig. 8. Decomposition of the Bragg curve by LET in water for the projectile-ion $^{12}C^{+6}$, 391 MeV/A.

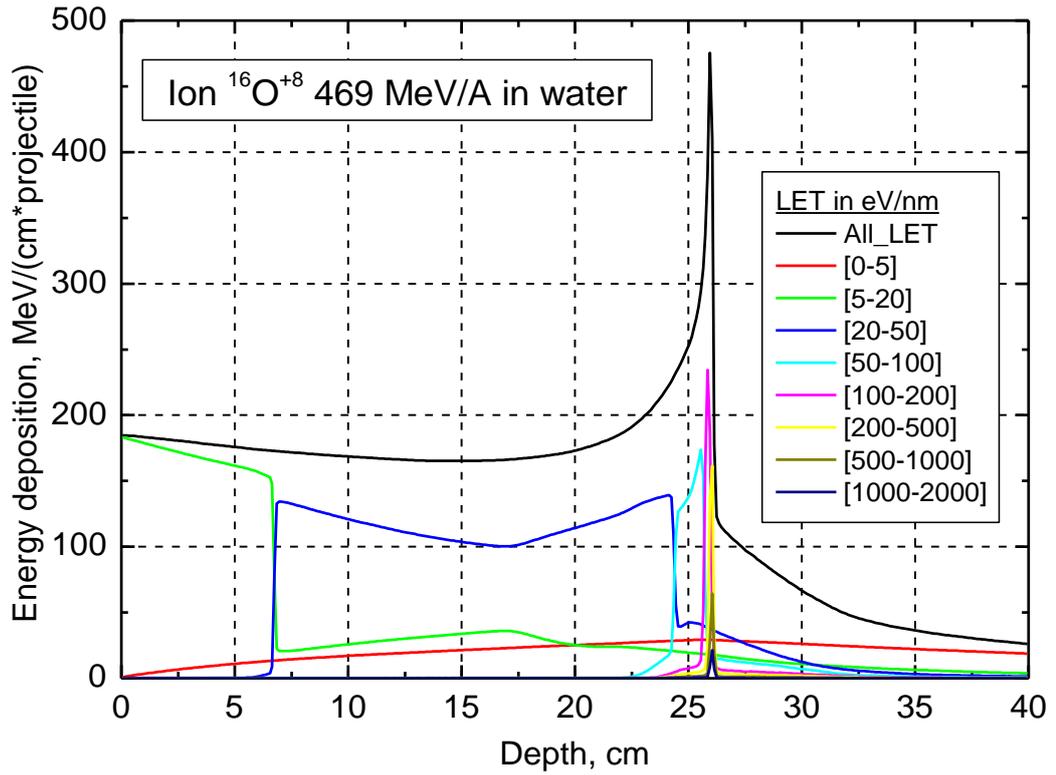

Fig. 9. Decomposition of the Bragg curve by LET in water for the projectile-ion $^{16}O^{+8}$, 469 MeV/A.

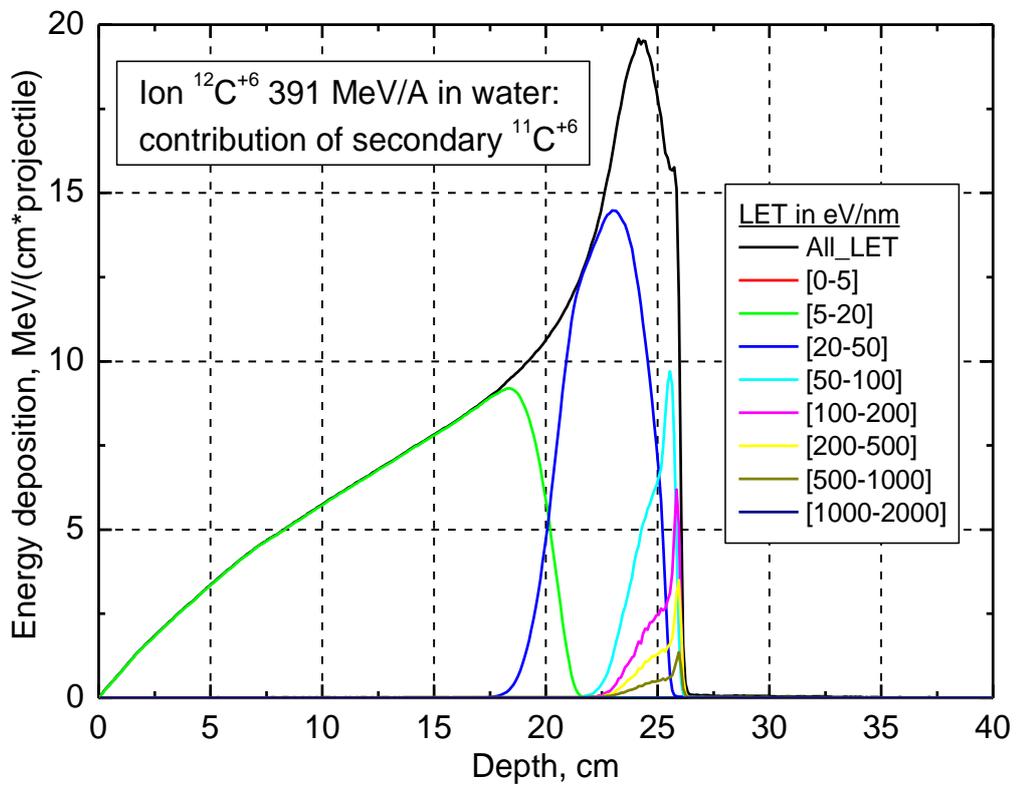

Fig. 10. Contribution of the secondary fragment $^{11}C^{+6}$ to the decomposition of Bragg curve by LET for the projectile-ion $^{12}C^{+6}$, at 391 MeV/A in water.

## 4. Conclusion

The described above option of the SHIELD-HIT code allows to decompose the absorbed dose by LET in tissue-equivalent materials, irradiated by beams of light ions in the therapeutic range of energies. This decomposition is possible for both the total energy deposition in the target, and of individual contributions of specific nuclear fragments. As an illustration the decomposition of the Bragg curves in water phantom at irradiation by protons, alpha particles, and by ions of lithium, carbon and oxygen is presented.

This option allows to analyze the absorbed dose as a function of LET in any volume inside a phantom. For example the presented results show (see Figs. 5-9) that the energy deposition behind the Bragg peak due to nuclear fragmentation occurs at more low LET and hence less dangerous for a healthy tissue. In fact the described option of the SHIELD-HIT code allows to build directly the dosimetric LET spectra. Since the measurements of LET spectra are labor-consuming, especially if the spectra of separate secondary fragments are measured [10], the computational approach is desirable. The model of nuclear fragmentation, which is used in the SHIELD-HIT code, was improved and well benchmarked in previous studies against experimental data and other Monte Carlo codes (see e.g. [4-6] and references therein).

**Acknowledgments**

This work is supported by the grant RFBR 15-52-46004 CN_a. TUBITAK support with project number 114F500 is gratefully acknowledged.